\documentclass[review,5p,fleqn]{elsarticle}

\usepackage[T1]{fontenc}
\usepackage{epsfig}
\usepackage{graphicx}
\usepackage{amsmath}
\usepackage{amsfonts}
\usepackage{amssymb}

\journal{Journal of Computational and Theoretical Chemistry}
\begin{document}
\begin{frontmatter}
\title{Construction and
counting of the number operators of an $n$-degree-of-freedom
normalized non-resonant vibrational  Hamiltonian}

\author[Dijon]{G.Saget}
\author[Dijon,Tomsk]{C. Leroy \corref{cor}}
\cortext[cor]{Corresponding author.}
\ead{claude.leroy@u-bourgogne.fr}
\author[Dijon]{H.R. Jauslin}
\address[Dijon]{Laboratoire Interdisciplinaire Carnot de Bourgogne (ICB),
UMR 6303 CNRS-Universit\'{e} Bourgogne Franche-Comt\'{e} 9 Av. A.
Savary, BP 47 870, F-21078 DIJON Cedex, FRANCE.}
\address[Tomsk]{Tomsk Polytechnic University, Institute of Physics and Technology, Department of General Physics, 30 Lenin Avenue, 634050 Tomsk, RUSSIA.}

\begin{abstract} \label{res}
The present paper is the first of two articles aimed at constructing
$n$-degree-of-freedom Hamiltonian systems  by an algebraic approach.
In molecular spectroscopy, the construction of vibrational
Hamiltonian for strongly excited molecular systems by using an
algebraic formalism  requires the introduction by hand the
 operators describing the change in energy by numerous
quanta and it is tedious to predict in advance the  total number of
operators appearing in the  development. The goal of the two
articles is not only to propose in the local limit a systematic
method for constructing a normalized vibrational  Hamiltonian for a
strongly excited $n$-degree-of-freedom molecular system from  the
generators of the Lie algebra, the algebra of polynomial invariants,
but also to  enumerate the number of independent operators needed
for the  construction of the Hamiltonian developed in the base of
these  generators up to the given order $N$. The first  article
introduces the theoretical tools used in the both papers (section
\ref{norm}), and presents the method of construction in case of
absence of resonance (section \ref{const}). Finally, an application
for a triatomic non-linear ClOH molecule is considered in case close
to the dissociation limit. (section \ref{Appli}).
\end{abstract}
\end{frontmatter}

\section{Introduction} \label{intro} Since years 1990, algebraic
formalisms are developed in molecular spectroscopy  (e.g.
\cite{IL1995}, \cite{KVB2013}). For instance, the vibrational
structure of local stretching modes for tetrahedral molecules
AB$_{4}$ has been studied through algebraic chains by
(\cite{LER1991}). To model molecular vibrations the theory of local
modes (e.g. \cite{CH1984, LM1992}) associates for each of $n$
degrees of freedom in vibration an oscillator  to each bond and to
each angular deformation. When molecular systems are strongly
excited or they are highly anharmonic already at small quantum
numbers, this description corresponds to a treatment that is more
adequate for vibrational modes than for the normal ones. \\
In classical mechanics  the study around a fixed
point  consists in writing the Hamiltonian $\mathcal{H}$
as a power series expansion of $\epsilon$
($\epsilon$ is a non-dimensional parameter  small compared to 1 and the
different Hamiltonians appearing in the development are of
the same order, and the development is up to the  order $\tilde{N}$, with
 $\tilde{N}$ being non-zero integer): $\mathcal{H} \approx \mathcal{H}_{0} +
\sum_{k=1}^{\tilde{N}} \epsilon^{k} \mathcal{H}_{k}$,
$\mathcal{H}_{0}$ being the Hamiltonian describing the system up to the
lowest order and $\mathcal{H}_{k}$ ($k=1,\,...,\tilde{N}$)
are the perturbation  Hamiltonians. It follows from this equation that
$\mathcal{H}_{0}$ is not rigorously the first integral
for the studied system. \\
Birkhoff (\cite{BIR1927}) then Gustavson (\cite{GUS1964}) or Deprit
((\cite{DEP1969}) proposed a so-called procedure of normalisation to
a new Hamiltonian $\mathcal{K}$ having a simpler form than
$\mathcal{H}$, called normalized Hamiltonian, and commuting with
 $\mathcal{H}_{0}$ in the sense of Poisson irregardless of the order
up to which it is developed. Just like $\mathcal{H}$, the normalized Hamiltonian
can be written as a power series development of
$\epsilon$; for the  development up to the order $\tilde{N}$, we have:
$\mathcal{K} \approx \mathcal{H}_{0} + \sum_{k=1}^{\tilde{N}}
\epsilon^{k}
\mathcal{K}_{k}$. \\
In this article, we  consider  the so-called standard normalisation
(\cite{CB1997}) of $n$-degree-of-freedom harmonic oscillator, i.e.
the Hamiltonian of the lowest order $\mathcal{H}_{0}$ is a
polynomial of the second order in  dynamic variables and reads:
$\mathcal{H}_{0} =
\sum_{k=1}^{n}\frac{\omega_{k}(q_{k}^{2}+p_{k}^{2})}{2}$, with
$\omega_{k}$ being the frequencies ($ 1 \leq i,\,j \leq n$,
$\omega_{i} \neq \omega_{j}$).

\section{Normalisation}\label{norm} \subsection{Hamiltonian of the
lowest order: $\mathcal{H}_{0}$}\label{ham} Let
$\mathcal{H}(q_{1},\,...,\, q_{n},\, p_{1},\,..., \, p_{n})$ be the
classical vibrational Hamiltonian  of an $n$-degree-of-freedom
Hamiltonian  system
 with the quadratic part being
that of an anisotropic harmonic oscillator
$\mathcal{H}_{0} =
\sum_{k=1}^{n}\frac{\omega_{k}(q_{k}^{2}+p_{k}^{2})}{2}$, where
$\omega_{k}$ are frequencies ($1 \leq
i,\,j \leq n$, $\omega_{i} \neq \omega_{j}$) and $q_{k}$ and
$p_{k}$ are the non-dimensional canonical variables  describing
the generalized coordinates  and their conjugated momenta.

\subsection{Equation of  motion}\label{Equations}
We introduce here
the complex variables $z_{k}$ and $z_{k}^{\ast}$ defined as
functions of canonical variables $q_{k}$ and $p_{k}$ ($k\in
[1,\,...,\,n]$) $z_{k}  = \frac{1}{\sqrt{2}}(q_{k}+ip_{k})$ and
$z_{k}^{\ast}  = \frac{1}{\sqrt{2}}(q_{k}-ip_{k})$. These variables
satisfy the relations: $1 \leq j,\,k \leq n$, $\{ z_{j},
z_{k}^{\ast} \} = \sum_{i=1}^{n}(\frac{\partial z_{j}}{\partial
q_{i}}\frac{\partial z_{k}^{\ast}}{\partial p_{i}}-\frac{\partial
z_{j}}{\partial p_{i}}\frac{ \partial z_{k}^{\ast}}{\partial
q_{i}})= -i\delta_{jk}$. The change of coordinates from
$(q_{k},\,p_{k})$ to $(z_{k},\,z_{k}^{\ast})$ is a  symplectic
transformation  of multiplier $-\imath$. The Hamiltonian
$\mathcal{H}_{0}$ reads: $\mathcal{H}_{0}=
-\imath\sum_{k=1}^{n}\omega_{k}z_{k} z_{k}^{\ast}$. In complexes
coordinates, for $1 \leq k \leq n$, the equations  of motion are:
\begin{equation}\label{eq.2.2.1} \frac{dz_{k}}{dt}  = \frac{\partial
\mathcal{H}_{0}}{\partial z_{k}^{\ast}} = -\imath \omega_{k} z_{k}.
\end{equation}

\subsection{Hamiltonian flow}\label{Flow} For an initial condition
$z_{0}=(z_{1,0},\,...,\,z_{n,0})$, the solution to the equation of
motion is written formally as $z(t)=
{\phi_{t}}^{\mathcal{H}_{0}}(z_{0})$. The application
${\phi_{t}}^{\mathcal{H}_{0}}:\Gamma \rightarrow \Gamma$ is the
Hamiltonian flow generated (\cite{ARN1976, EFS2005}) by
$\mathcal{H}_{0}$. \\ We have: $z(t) =
{\phi_{t}}^{\mathcal{H}_{0}}(z_{0}) \Longrightarrow \left(
                                                    \begin{array}{c}
                                                      z_{1}(t) \\
                                                      \vdots \\
                                                      z_{n}(t) \\
                                                    \end{array}
                                                  \right)$
\begin{eqnarray}\label{eq.2.3.1}
=\left(
   \begin{array}{cccc}
     e^{-\imath \omega_{1}t} & 0 & \ldots & 0  \\
     0 & e^{-\imath \omega_{2}t} &  \ldots & 0 \\
     0 & 0 & \ddots & 0 \\
     0 & 0 & 0 & e^{-\imath \omega_{n}t}
   \end{array}
 \right)
 \left(
                                              \begin{array}{c}
                                                z_{1,0} \\
                                                \vdots \\
                                                z_{n,0} \\
                                              \end{array}
                                            \right).
\end{eqnarray} The  solutions $z(t)$ still constitute   the
trajectories of the Hamiltonian flow or the orbits of the harmonic
oscillator.

\subsection{Normalisation tools}\label{tools}
\subsubsection{Mathematical condition of non-resonance}
The Hamiltonian $\mathcal{H}_{0}$ is called non-resonant (for example
\cite{MHO2009}) if the $n$ frequencies $\omega_{k}$ are
independents, i.e. ($\lambda_{k}$ are all zero integers):
\begin{eqnarray}\label{eq.2.4.1}
\sum_{k=1}^{n} \lambda_{k}\omega_{k} = 0.
\end{eqnarray}

\subsubsection{Adjoint operator $ad_{\mathcal{H}_{0}}$}
The adjoint operator associated to Hamiltonian $\mathcal{H}_{0}$ is
 the linear operator acting in the phase space
$\Gamma$ (\cite{BHLV2003}) \\
$ad_{\mathcal{H}_{0}}$: $\forall$ $\mathcal{G} \in \Gamma$,
$\mathcal{G}$
 $\rightarrow$ $ad_{\mathcal{H}_{0}}(\mathcal{G})$ $=$ $\{\mathcal{G}, \mathcal{H}_{0} \} = \sum_{i=1}^{n}
\biggl(\frac{\partial \mathcal{G}}{\partial q_{i}}\frac{\partial
\mathcal{H}_{0}}{\partial p_{i}} - \frac{\partial
\mathcal{G}}{\partial p_{i}}\frac{\partial \mathcal{H}_{0}}{\partial
q_{i}}\biggr)$. In complex coordinates, this operator reads:
\begin{equation}\label{eq.2.4.2} ad_{\mathcal{H}_{0}} =
\sum_{k=1}^{n}\omega_{k}\biggl(z_{k}^{\ast}\frac{\partial}{\partial{z_{k}^{\ast}}}-z_{k}\frac{\partial}{\partial{z_{k}}}\biggr).
\end{equation}

\subsubsection{Sets $Ker$ $ad_{\mathcal{H}_{0}}$ and $Im$
$ad_{\mathcal{H}_{0}}$} The  complex monomials  $\sigma =
{z_{1}}^{\alpha_{1}}{{z_{1}}^{\ast}}^{\beta_{1}}...{z_{n}}^{\alpha_{n}}{{z_{n}}^{\ast}}^{\beta_{n}}$
(with $\sum_{k=1}^{n}(\alpha_{k}+\beta_{k}) = m$ being the monomial
degree, and $\alpha_{k}$ and $\beta_{k}$ being integers) are the
elements of the complex polynomials circle
$\mathbb{C}(z,\,z^{\ast})$ and are the eigenvectors of
$ad_{\mathcal{H}_{0}}$ as: \begin{equation}\label{eq.2.4.3}
ad_{\mathcal{H}_{0}}(\sigma)=
\biggl(\sum_{k=1}^{n}\omega_{k}(\beta_{k}-\alpha_{k})\biggr)(\sigma).
\end{equation} $Ker$ $ad_{\mathcal{H}_{0}}$ (kernel of
$ad_{\mathcal{H}_{0}}$) and $Im$ $ad_{\mathcal{H}_{0}}$ (image of
$ad_{\mathcal{H}_{0}}$) verify (resp.)
$ad_{\mathcal{H}_{0}}(\sigma)=0$ and $ad_{\mathcal{H}_{0}}(\sigma)
\neq 0$.

\subsubsection{Circles of complex polynomials  invariant with the
time reversal} The application of the time reversal operator $\tau$
on the set $(z_{k},\,{z_{k}}^{\ast})$ ($1\leq k \leq n$) gives:
$\tau(z_{k})={z_{k}}^{\ast}$, $\tau({z_{k}}^{\ast})=z_{k}$. We
denote by $\mathbb{C}(z,\,z^{\ast})^{\tau}$ the sub-circle of the
monomials $\sigma$ invariant with respect to the time reversal
(\cite{STU1993}): $\mathbb{C}(z,\,z^{\ast})^{\tau}=\{\sigma \in
\mathbb{C}(z,\,z^{\ast}) / \tau(\sigma)=\sigma\}$.

\subsubsection{Hilbert base}
Due to the condition of non-resonance (\ref{eq.2.4.1}), the kernel
of  adjoint operator $ad_{\mathcal{H}_{0}}$ is produced (created) by the
$n$ monomials $\sigma_{1} = {z_{1}}{z_{1}}^{\ast}$, ..., $\sigma_{n} =
{z_{n}}{z_{n}}^{\ast}$ all belonging to
$\mathbb{C}(z,\,z^{\ast})^{\tau}$. $Ker$ $ad_{\mathcal{H}_{0}}$ has a
  a Lie algebra structure called algebra of
invariant polynomials. The generators $\sigma_{k}$ ($1\leq k \leq
n$) form a base of $Ker$ $ad_{\mathcal{H}_{0}}$ called Hilbert base
(\cite{HIL1893, GAT2000, CLO1992}). This base is unique for a
non-resonant Hamiltonian (\cite{BHLV2003}).\\ The  generators are
invariants with respect to  flow of the harmonic oscillator, i.e.:
$\forall t,\, 1 \leq k \leq n, \,\forall z \in \Gamma$,
\begin{eqnarray}\label{eq.2.4.6} \sigma_{k}
(\phi_{t}^{\mathcal{H}_{0}}(z))= \sigma_{k}(z)\,. \end{eqnarray}
(\ref{eq.2.4.6}) means that $\phi_{t}^{\mathcal{H}_{0}}$ is a
symplectic symmetry  for the  generators of the  algebra
(\cite{ARN1976, EFS2005, MHO2009}); in other terms, according to the
Noether theorem, $1 \leq k \leq
n,\,\,\{\sigma_{k},\,\mathcal{H}_{0}\}=0$. \\ Reciprocally, we know
that by construction the generators $\sigma_{k}$ belong to $Ker$
$ad_{\mathcal{H}_{0}}$, so $1 \leq k \leq n$, $\{\sigma_{k},
\mathcal{H}_{0}\} = 0$. If $\mathcal{H}_{0}$ is an invariant for the
generators then, according to the  "reciprocality" of the Noether
theorem, $\phi_{t}^{\mathcal{H}_{0}}$ is a symplectic symmetry  for
the generators. Thus we have : \begin{eqnarray}\label{eq.2.4.7}
\forall t,\, 1 \leq k \leq n, \,\forall z\in \Gamma,\,\,\sigma_{k}
(\phi_{t}^{\mathcal{H}_{0}}(z))= \sigma_{k}(z), \notag \\
\Longleftrightarrow \,\, \{\sigma_{k}, \mathcal{H}_{0}\} = 0\,.
\end{eqnarray} In  complex coordinates, the  Poisson brackets
between the generators read (pour $1 \leq i,\,j\leq n$):
\begin{eqnarray}\label{eq.2.4.8} \{\sigma_{i}, \sigma_{j}\}  =
i\sum_{k=1}^{n}(\frac{\partial
\sigma_{i}}{\partial{z_{k}}^{\ast}}\frac{\partial
\sigma_{j}}{\partial{z_{k}}}-\frac{\partial
\sigma_{i}}{\partial{z_{k}}}\frac{\partial
\sigma_{j}}{\partial{z_{k}}^{\ast}}) = 0\,. \end{eqnarray}

\subsubsection{Normalized Hamiltonian  $\mathcal{K}$}
We look for constructing a Hamiltonian called normalized Hamiltonian
$\mathcal{K}$ such that $\mathcal{H}_{0}$ commute with $\mathcal{K}$,
i.e. $\{\mathcal{H}_{0},\,\mathcal{K}\} =
-ad_{\mathcal{H}_{0}}(\mathcal{K})=ad_{\mathcal{K}}(\mathcal{H}_{0})=0$.
The normalized Hamiltonian has the following form (\cite{BHLV2003}):
$\mathcal{K} = \mathcal{H}_{0} + f(\sigma_{1},\,...,\, \sigma_{n})$.
As  the generators of the Hilbert base are invariant with respect to
the time reversal, the Hamiltonian is invariant as well:
$\tau(\mathcal{K})= \mathcal{K}$.

\subsubsection{Order of the  development}
In two articles, we  write the normalized Hamiltonian
as a  power polynomial development of the generators.
In what follows, we denote by $\lambda$-monomial of degree
$d=\sum_{\ell=1}^{\lambda}2r_{i_{\ell}}$ each monomial formed by
the product of the powers of $\lambda$ generators of the Hilbert base
 ($2\leq \lambda \leq n$):
${\sigma_{i_{1}}}^{r_{i_{1}}}...{\sigma_{i_{\lambda}}}^{r_{i_{\lambda}}}$
(obviously $d \geq 2\lambda$). The order $N$ of the  development is
fixed by the $\lambda$-monomials of the highest
 degree $d=N$ of the Hamiltonian
$\mathcal{K}$. In the first article as the generators of the Hilbert
base are all of the order 2, the order $N$ is necessarily pair (with
$N\geq 2$).

\section{Construction of counting}\label{const}
\subsection{Case $n=2$}
\subsubsection{Construction of the normalized Hamiltonian }
This case can model, for example, the two stretching local modes of
rigid triatomic linear molecules ABC, and  also all  molecular
systems having at least two vibrational non-resonant modes. \\ The
normalized non-resonant Hamiltonian $\mathcal{K}$, with the
quadratic part $\mathcal{H}_{0} =
-\imath(\omega_{1}\sigma_{1}+\omega_{2} \sigma_{2})$, is only the
function of $\sigma_{1}$ et $\sigma_{2}$: $\mathcal{K} =
\mathcal{H}_{0} + f(\sigma_{1},\, \sigma_{2})$. \\ Let us write it
down as a polynomial development  of $\sigma_{1}$ and $\sigma_{2}$
up to the order $N$ ($N \geq 4$ is natural integer):
\begin{eqnarray}\label{eq.3.1.1} \mathcal{K} & = & \mathcal{H}_{0} +
\sum_{q_{0} = 2}^{Q_{0}}({\alpha_{q_{0}}^{1}} {\sigma_{1}}^{q_{0}} +
{\alpha_{q_{0}}^{2}} {\sigma_{2}}^{q_{0}}) \notag \\ &+&
\sum_{r=2}^{Q_{0}}\sum_{1\leq i_{1} < i_{2} \leq 2}\sum_{r_{i_{1}}
\geq 1,\,r_{i_{2}}\geq1}^{r_{i_{1}}+r_{i_{2}} = r}
{\alpha_{r_{i_{1}},\,r_{i_{2}}}^{i_{1},\,i_{2}}}
{\sigma_{i_{1}}^{r_{i_{1}}}}{\sigma_{i_{2}}^{r_{i_{2}}}}.
\end{eqnarray} In Eq. (\ref{eq.3.1.1}), the monomials are all real.
Moreover, the transformation of $(q_{k},\,p_{k})$ to
$(z_{k},\,z_{k}^{\ast})$ is a symplectic transformation  of
multiplier $-i$, in the manner of the Hamiltonian $\mathcal{K}$, we
deduce that all the coefficients ${\alpha_{q_{0}}^{1}}$,
${\alpha_{q_{0}}^{2}}$,
${\alpha_{r_{i_{1}},\,r_{i_{2}}}^{i_{1},\,i_{2}}}$ are pure
imaginary; $q_{0}$, $i_{1}$, $i_{2}$, $r_{i_{1}}$ and $r_{i_{2}}$
are natural integers; in (\ref{eq.3.1.1}), the degree $d$ of the
monomials $\sigma_{1}^{q_{0}}$, $\sigma_{2}^{q_{0}}$ and
${\sigma_{i_{1}}^{r_{i_{1}}}}{\sigma_{i_{2}}^{r_{i_{2}}}}$ is equal
to $2q_{0}$ and $2r$, respectively; $q_{0}$ and $r$ take all the
integer values from $2$ to a maximal value  $Q_{0}$ such that $Q_{0}
= E(\frac{N}{2})$ with $E(x)$ being the integer (entire) part of
$x$. If $N =2$, the coefficients
${\alpha_{r_{i_{1}},\,r_{i_{2}}}^{i_{1},\,i_{2}}}$,
${\alpha_{q_{0}}^{1}}$ and ${\alpha_{q_{0}}^{2}}$ are zero.
Moreover, $\mathcal{K}$ contains $\frac{Q_{0}(Q_{0}+3)}{2}$
monomials.

\subsubsection{Independence of the  coefficients}\label{ind}
Let us write down the Jacobi equality  $(1\leq j \leq 2)$:
$\{\mathcal{K},\,\{\sigma_{j},\,\mathcal{H}_{0}\}\}$
$+\{\mathcal{H}_{0},$ $\,\{\mathcal{K},\,\sigma_{j}\}\}+
\{\sigma_{j},\,\{\mathcal{H}_{0},\,\mathcal{K}\}\}=0$. Calculating
the different Poisson brackets $\{\mathcal{K},\,\sigma_{j}\}$ and
knowing that $(\mathcal{K},\,\sigma_{j}) \in Ker$
$ad_{\mathcal{H}_{0}}$, we deduce that there is no relation
between the coefficients of the development of  Eq. (\ref{eq.3.1.1}):
the different $\lambda$-monomials are independent between them. The
Hamiltonian is  described by as many  monomials as the coefficients.

\subsection{Case $n=3$} This mathematical study  can be used as a
model, for example, for rigid molecules to describe local modes
(stretching or bending) of triatomic non linear molecules, ABC or
AB$_{2}$. \\ The normalized non-resonant Hamiltonian $\mathcal{K}$
with the quadratic part $\mathcal{H}_{0} =
-\imath(\omega_{1}\sigma_{1}+\omega_{2} \sigma_{2}+\omega_{3}
\sigma_{3})$ is only the function of $\sigma_{1}$, $\sigma_{2}$ and
$\sigma_{3}$: $\mathcal{K} = \mathcal{H}_{0} + f(\sigma_{1},\,
\sigma_{2},\, \sigma_{3})$. \\
Let us write down it as a polynomial development of $\sigma_{1}$,
$\sigma_{2}$ and $\sigma_{3}$ up to the order $N$:

\begin{eqnarray*}
\mathcal{K} & = & \mathcal{H}_{0} + \sum_{q_{0} =
2}^{Q_{0}}({\alpha_{q_{0}}^{1}} {\sigma_{1}}^{q_{0}} +
{\alpha_{q_{0}}^{2}} {\sigma_{2}}^{q_{0}} +  {\alpha_{q_{0}}^{3}}
{\sigma_{3}}^{q_{0}})+
\end{eqnarray*}
\begin{eqnarray}\label{eq.3.2.1}
\sum_{\ell=2}^{3}\sum_{r=2}^{Q_{0}}\sum_{1\leq i_{1} <\ldots<
i_{\ell} \leq 3}\sum_{r_{i_{1}} \geq
1,\,\ldots,\,r_{i_{\ell}}\geq1}^{r_{i_{1}}+\ldots +r_{i_{\ell}} = r}
{\alpha_{r_{i_{1}},\,...,\,r_{i_{\ell}}}^{i_{1},\,...,\,i_{\ell}}}{\sigma_{i_{1}}^{r_{i_{1}}}}\ldots{\sigma_{i_{\ell}}^{r_{i_{\ell}}}}.
\end{eqnarray}
$i_{1}$, ..., $i_{\ell}$ are non-zero natural integers
satisfying the order relations: $1\leq i_{1} < \ldots < i_{\ell}
\leq 3$; $r_{i_{1}}$, ..., $r_{i_{\ell}}$ are non-zero natural integers satisfying
 the relation $r_{i_{1}}+...+r_{i_{\ell}} =
r$, with $r$ being the natural integer  comprised between $2$ and
$Q_{0}$. The coefficients ${\alpha_{q_{0}}^{k}}$ ($1 \leq k \leq
3$),
${\alpha_{r_{i_{1}},\,...,\,r_{i_{\ell}}}^{i_{1},\,...,\,i_{\ell}}}$
are the pure imaginary parameters.  For $N<6$ all the terms of the
last summation of the rhs of Eq. (\ref{eq.3.2.1}) are identically
equal to zero. Besides, $\mathcal{K}$ contains
$\frac{Q_{0}(Q_{0}^{2}+Q_{0}+11)}{6}$ monomials.

\subsection{General case}
\subsubsection{Construction of the normalized  Hamiltonian }
The quadratic part of the normalized non-resonant Hamiltonian
$\mathcal{K}$ reads: $\mathcal{H}_{0} =
-\imath\sum_{k=1}^{n}\omega_{k}\sigma_{k}$. The $n$ quantities
$\omega_{k}$ are the caracteristic frequencies
of the oscillators. \\
Let us write $\mathcal{K}$ as a function of $n$ generators of the
 Hilbert base: $\mathcal{K} = \mathcal{H}_{0} +
f(\sigma_{1},\,..., \, \sigma_{n})$. \\
In case of absence of resonance between  two any oscillators,
let us write down $\mathcal{K}$ as a polynomial development
(development of Dunham) of $\sigma_{1}$, $\sigma_{2}$, ...,
$\sigma_{n}$ (with $n \geq 3$ being natural integer) up to the order $N$:
\begin{eqnarray*}
\mathcal{K}  =  \mathcal{H}_{0} +
\sum_{k=1}^{n}\sum_{q_{0}=2}^{Q_{0}}
{\alpha_{q_{0}}^{k}}{\sigma_{k}^{q_{0}}}+
\end{eqnarray*}
\begin{eqnarray}\label{eq.3.3.1}
\sum_{\ell=2}^{n}\sum_{r=2}^{Q_{0}}\sum_{1\leq i_{1} < i_{2} <
\ldots < i_{\ell} \leq n}\sum_{r_{i_{1}} \geq
1,\,\ldots,\,r_{i_{\ell}}\geq1}^{r_{i_{1}}+\ldots +r_{i_{\ell}} = r}
{\alpha_{r_{i_{1}},\,...,\,r_{i_{\ell}}}^{i_{1},\,...,\,i_{\ell}}}
{\sigma_{i_{1}}^{r_{i_{1}}}}\ldots{\sigma_{i_{\ell}}^{r_{i_{\ell}}}}.
\end{eqnarray}
$i_{1}$, ..., $i_{\ell}$ are the non-zero natural integers
satisfying the order relations : $1\leq i_{1} < i_{2} < \ldots <
i_{\ell} \leq n$; $r_{i_{1}}$, ..., $r_{i_{\ell}}$ are the non-zero natural integers
satisfying a relation
$r_{i_{1}}+...+r_{i_{\ell}} = r$, with $r$ being a natural integer  comprised
between $2$ and $Q_{0}$. The coefficients ${\alpha_{q_{0}}^{k}}$ and
${\alpha_{r_{i_{1}},\,...,\,r_{i_{\ell}}}^{i_{1},\,...,\,i_{\ell}}}$
are the pure imaginary parameters.  For $N<6$, all the terms of
the last summation of the rhs of Eq. (\ref{eq.3.3.1}) are
identically equal to zero. Proceeding as in Section
\ref{ind}, we verify that  all the $\lambda$-monomials in Eq.
(\ref{eq.3.3.1}) are independent.

\subsubsection{Counting}
The  counting of the  number $\Lambda$ of independent operators
required to construct $\mathcal{K}$ for the case $n=3$
can be presented in the form  ($N \geq 6$):
\begin{eqnarray}\label{eq.3.3.2}
\Lambda & = & 3+ 3(Q_{0}-1)+\frac{3Q_{0}(Q_{0}-1)}{2} \notag \\
&+& \frac{Q_{0}(Q_{0}-1)(Q_{0}-2)}{6}, \notag \\
\Lambda  & = &
\sum_{\lambda=1}^{3}{C_{3}^{\lambda}}{C_{Q_{0}}^{\lambda}}.
\end{eqnarray}
Let us generalize the counting formula (\ref{eq.3.3.2}) for an
$n$-degree-of-freedom system ($n\geq 3$). \\
For the order  $N$ and $d$ data ($2\lambda \leq d \leq N$), there is
$C_{n}^{\lambda}$ ways of choosing the $\lambda$ generators  of a
$\lambda$-monomial of degree $d$ among the set ($\sigma_{1}$,
$\sigma_{2}$, $\sigma_{3}$, ..., $\sigma_{n}$). Moreover, for a given
choice of $\lambda$ generators, let us denote by $N_{r}$ the number of ways
to realize a $\lambda$-monomial
${\sigma_{i_{1}}^{r_{i_{1}}}}\ldots{\sigma_{i_{\lambda}}^{r_{i_{\lambda}}}}$,
with the set of powers $(r_{i_{1}},\,\ldots,\,r_{i_{\lambda}})$
that should satisfy, in accordance to Eq. (\ref{eq.3.3.1}), the condition
$r_{i_{1}}+\ldots+r_{i_{\lambda}}=r$. Note that
$C_{Q_{0}}^{\lambda}=\sum_{r=\lambda}^{Q_{0}}N_{r}$. We deduce from this that
for the given $N$     and $\lambda$, there is
$C_{n}^{\lambda}C_{Q_{0}}^{\lambda}$ independent monomials entering
the composition of $\mathcal{K}$. Finally, as the maximal value
of $\lambda$ is  fixed either by the number of the degree of freedom
of the Hamiltonian system  if $Q_{0} \geq n$, or either by $Q_{0}$
if $Q_{0} \leq n$, the total number of independent monomials in
$\mathcal{K}$ is obtained by adding the
$C_{n}^{\lambda}C_{Q_{0}}^{\lambda}$ of $\lambda=1$ to the smallest value of
$n$ or $Q_{0}$: $min(n,\,Q_{0})$. It follows that
$\mathcal{K}$ contains:
\begin{equation}\label{eq.3.3.4}
\Lambda =
\sum_{\lambda=1}^{min(n,\,Q_{0})}{C_{n}^{\lambda}}{C_{Q_{0}}^{\lambda}}
\,\,\text{monomials}\,.
\end{equation}

As an example, Table \ref{table.1} shows, for $n=3$ and $n=4$
the number of monomials $\Lambda$ in $\mathcal{K}$ developed up to the order $N=8$, as
well as the list of monomials. \\

\begin{table}[h]
\begin{center}
\begin{eqnarray*}
\begin{array}{|c|c|c|} \hline
n &  \Lambda & \text{monomials} \\
\hline 3 & 34 & \sigma_{1},\,\sigma_{2},\,\sigma_{3} \\
         &    & \sigma_{1}^{2},\,\sigma_{2}^{2},\,\sigma_{3}^{2},\,\sigma_{1}^{3},\,\sigma_{2}^{3},\,\sigma_{3}^{3},\,\sigma_{1}^{4},\,\sigma_{2}^{4},\,\sigma_{3}^{4}\\
         &    &  \sigma_{1}\sigma_{2},\,{\sigma_{1}}^{2}\sigma_{2},\,\sigma_{1}{\sigma_{2}}^{2},\,{\sigma_{1}}^{3}\sigma_{2},\,{\sigma_{1}}^{2} {\sigma_{2}}^{2},\,\sigma_{1}{\sigma_{2}}^{3} \\
         &    & \sigma_{1}\sigma_{3},\,{\sigma_{1}}^{2} \sigma_{3},\,\sigma_{1}{\sigma_{3}}^{2},\,{\sigma_{1}}^{3}\sigma_{3},\,{\sigma_{1}}^{2}{\sigma_{3}}^{2},\,\sigma_{1}{\sigma_{3}}^{3} \\
         &    & \sigma_{2}\sigma_{3},\,{\sigma_{2}}^{2}\sigma_{3},\,\sigma_{2}{\sigma_{3}}^{2},\,{\sigma_{2}}^{3}\sigma_{3},\,{\sigma_{2}}^{2}{\sigma_{3}}^{2},\,\sigma_{2}{\sigma_{3}}^{3} \\
         &    &  \sigma_{1}\sigma_{2}\sigma_{3},\,{\sigma_{1}}^{2}\sigma_{2}\sigma_{3},\,\sigma_{1}{\sigma_{2}}^{2}\sigma_{3},\,\sigma_{1}\sigma_{2}{\sigma_{3}}^{2}
         \\
\hline 4 & 69 & \sigma_{1},\,\sigma_{2},\,\sigma_{3},\,\sigma_{4} \\
         &    &  \sigma_{1}^{2},\,\sigma_{2}^{2},\,\sigma_{3}^{2},\,\sigma_{4}^{2},\,\sigma_{1}^{3},\,\sigma_{2}^{3},\,\sigma_{3}^{3},\,\sigma_{4}^{3},\,\sigma_{1}^{4},\,\sigma_{2}^{4},\,\sigma_{3}^{4},\,\sigma_{4}^{4}\\
         &    & \sigma_{1}\sigma_{2},\,{\sigma_{1}}^{2}\sigma_{2},\,\sigma_{1}{\sigma_{2}}^{2},\,{\sigma_{1}}^{3}\sigma_{2},\,{\sigma_{1}}^{2}{\sigma_{2}}^{2},\,\sigma_{1}{\sigma_{2}}^{3} \\
         &    & \sigma_{1} \sigma_{3},\,{\sigma_{1}}^{2}\sigma_{3},\,\sigma_{1}{\sigma_{3}}^{2},\,{\sigma_{1}}^{3} \sigma_{3},\,{\sigma_{1}}^{2} {\sigma_{3}}^{2},\,\sigma_{1}{\sigma_{3}}^{3} \\
         &    & \sigma_{1} \sigma_{4},\,{\sigma_{1}}^{2}\sigma_{4},\,\sigma_{1}{\sigma_{4}}^{2},\,{\sigma_{1}}^{3} \sigma_{4},\,{\sigma_{1}}^{2} {\sigma_{4}}^{2},\,\sigma_{1}{\sigma_{4}}^{3} \\
         &    & \sigma_{2} \sigma_{3},\,{\sigma_{2}}^{2}\sigma_{3},\,\sigma_{2}{\sigma_{3}}^{2},\,{\sigma_{2}}^{3} \sigma_{3},\,{\sigma_{2}}^{2} {\sigma_{3}}^{2},\,\sigma_{2}{\sigma_{3}}^{3} \\
         &    & \sigma_{2} \sigma_{4},\,{\sigma_{2}}^{2}\sigma_{4},\,\sigma_{2}{\sigma_{4}}^{2},\,{\sigma_{2}}^{3} \sigma_{4},\,{\sigma_{2}}^{2} {\sigma_{4}}^{2},\,\sigma_{2}{\sigma_{4}}^{3} \\
         &    & \sigma_{3} \sigma_{4},\,{\sigma_{3}}^{2}\sigma_{4},\,\sigma_{3}{\sigma_{4}}^{2},\,{\sigma_{3}}^{3} \sigma_{4},\,{\sigma_{3}}^{2} {\sigma_{4}}^{2},\,\sigma_{3}{\sigma_{4}}^{3} \\
         &    & \sigma_{1}\sigma_{2}\sigma_{3},\,{\sigma_{1}}^{2}\sigma_{2}\sigma_{3},\,\sigma_{1}{\sigma_{2}}^{2}\sigma_{3},\,\sigma_{1}\sigma_{2}{\sigma_{3}}^{2} \\
         &    & \sigma_{1}\sigma_{2}\sigma_{4},\,{\sigma_{1}}^{2}\sigma_{2}\sigma_{4},\,\sigma_{1}{\sigma_{2}}^{2}\sigma_{4},\,\sigma_{1}\sigma_{2}{\sigma_{4}}^{2} \\
         &    & \sigma_{1}\sigma_{3}\sigma_{4},\,{\sigma_{1}}^{2}\sigma_{3}\sigma_{4},\,\sigma_{1}{\sigma_{3}}^{2}\sigma_{4},\,\sigma_{1}\sigma_{3}{\sigma_{4}}^{2} \\
         &    & \sigma_{2}\sigma_{3}\sigma_{4},\,{\sigma_{2}}^{2}\sigma_{3}\sigma_{4},\,\sigma_{2}{\sigma_{3}}^{2}\sigma_{4},\,\sigma_{2}\sigma_{3}{\sigma_{4}}^{2} \\
         &    & \sigma_{1}\sigma_{2}\sigma_{3}\sigma_{4} \\
\hline
\end{array}
\end{eqnarray*}
\caption{Counting and list of monomials appearing in the
normalized Hamiltonian  $\mathcal{K}$ up to the order $N=8$ for $n=3$
et $n=4$.}\label{table.1}
\end{center}
\end{table}
The case $n=4$ can describe, for rigid molecules, two stretching non-resonant local modes and
 two non-resonant bending modes of  tetrahedral  molecules AB$_{3}$C.

\section{Application}\label{Appli} \subsection{Molecule ClOH ($n=3$)}
\subsubsection{Conventions of notation}
The molecule ClOH is a triatomic non-linear molecule  of 3
vibrational degrees of freedom ($n=3$). In the local limit, we
attach a stretching oscillator  to bond Cl-O (oscillator "1") and
O-H (oscillator "3") and a bending oscillator
 to the angle between these bonds (oscillator "2").

\subsubsection{Quantum vibrational Hamiltonian}
The classical  relations
 between the non-dimensional variables  $\{z_{j},
\,z_{k}^{\ast}\} = -i\delta_{jk}$, are now replaced by
$\frac{1}{i}[\hat{a_{j}}, \hat{{a_{k}^{+}}}] = -i\delta_{jk}$, i.e.:
$1\leq j,\,k \leq n$, $[\hat{a_{j}}, \hat{{a_{k}^{+}}}] =
\delta_{jk}$. These operators satisfy the commutation Bose relations
and are the boson creation $\hat{{a_{k}^{+}}}$ and
annihilation $\hat{a_{j}}$ operators. \\
The generators $\sigma_{k}$ and the  Hamiltonian function
$\mathcal{K}$ are correspondingly replaced by the number operators
 $\hat{N}_{k}= \hat{{a_{k}^{+}}}\hat{a_{k}}$, which
represent from the physical point of view the number of excitation
quanta per oscillator, and Hamiltonian operator $\hat{K}$. In the
development of $\hat{K}$ all the operators are hermitic and the
coefficients are real.

\subsubsection{Eigenbase of the Hamiltonian $\hat{H_{0}}$} The
eigenstates of $\hat{H_{0}}$ are generated from the empty state by
the relation ($n_{1}$, $n_{2}$, $n_{3}$ are the natural integers):
\begin{equation}\label{eq.4.1.1} \left| \begin{array}{c} n_{1},
n_{2},n_{3} \\ \end{array} \right >
=(n_{1}!n_{2}!n_{3}!)^{-\frac{1}{2}}\
{\hat{{a_{1}^{+}}}}^{n_{1}}{\hat{{a_{2}^{+}}}}^{n_{2}}{\hat{{a_{3}^{+}}}}^{n_{3}}
\left| \begin{array}{c} 0,0,0 \\ \end{array} \right >\,.
\end{equation}

\subsubsection{Numeric simulations }
\begin{table}[h] \begin{center} \begin{eqnarray*}
\begin{array}{|c|c|c|} \hline N &  \Lambda & \text{coefficients
in cm}^{-1} \\ \hline 2 & 3 & \omega_{1}= +739.685, \,
\omega_{2}=+1\,245.09\\
 & & \omega_{3}= +3\,748.47 \\
\hline
4 & 9 & {\alpha_{2}^{1}} = -3.517, \,{\alpha_{2}^{2}}=+2.181,\, {\alpha_{2}^{3}}=-84.540 \\
  &   & \alpha_{1,1}^{1,2}=-7.049,\,\alpha_{1,1}^{1,3}= -0.490,\,\alpha_{1,1}^{2,3} = -16.291\\
\hline
6 & 19 & {\alpha_{3}^{1}}= -0.259, \, {\alpha_{3}^{2}}= -0.778,\, {\alpha_{3}^{3}}= -0.173  \\
 &   & \alpha_{1,2}^{1,2}=-0.131,\,\alpha_{1,2}^{1,3}= -0.122,\,\alpha_{1,2}^{2,3} = -3.965\\
 &   & \alpha_{2,1}^{1,2}=-0.428,\,\alpha_{2,1}^{1,3}= -0.508,\,\alpha_{2,1}^{2,3} = -0.154\\
 &   & \alpha_{1,1,1}^{1,2,3}=-0.767 \\
\hline
8 & 34 & {\alpha_{4}^{1}}= +0.0098 , \, {\alpha_{4}^{2}}= +0.0111,\, {\alpha_{4}^{3}}= +0.0153   \\
&   &
\alpha_{3,1}^{1,2}=0,\,\alpha_{3,1}^{1,3}=0,\,\alpha_{3,1}^{2,3}=+0.0793
\\
&   &
\alpha_{2,2}^{1,2}=-0.0079,\,\alpha_{2,2}^{1,3}=-0.0174,\,\alpha_{2,2}^{2,3}=-0.0426
\\
&   & \alpha_{1,3}^{1,2}=+0.0021,\,\alpha_{1,3}^{1,3}=-0.0007
\\
& &  \alpha_{1,3}^{2,3}=+0.2885
\\
&   & \alpha_{1,1,2}^{1,2,3}=
+0.1553,\,\alpha_{1,2,1}^{1,2,3}=+0.1003
\\
& & \alpha_{2,1,1}^{1,2,3}=+0.0854
\\
\hline
\end{array}
\end{eqnarray*}
\caption{Number of  Dunham parameters  given  by \cite{JJS1999}. For
the given order  $N$ ($4\leq N \leq 8$), a line contains the number
of additional operators with respect to the order
$N-2$.}\label{table.2}
\end{center}
\end{table}

The vibrational  structure of the molecule ClOH has been studied in
\cite{JJS1999} nearly up to the dissociation limit (the energies are
given relative to the $|0,0,0>$ ground state located at $2\,867.0$
cm$^{1}$ above the bottom of the potential energy surface. The
dissociation energy is $19 \,290$ cm$^{-1}$). The authors realize a
Dunham development   of number operators up to the order $N=8$; such
the  constructed Hamiltonian  allows one to reproduce the
vibrational structure of  314 energy levels up to $70$ $\%$ of the
dissociation  energy (i.e. $13 \,500$ cm$^{-1}$) requiring $34$
coefficients which is in accordance with Eq. (\ref{eq.3.3.1}) with
the coefficients (enumerated by Eq. (\ref{eq.3.3.4})) given in Table
\ref{table.2}.

\section{Conclusion and perspectives}\label{Concl}
The counting of number operators of an $n$-degree-of-freedom
vibrational non-resonant Hamiltonian, developed in the base of
number operators  $\hat{N}_{k}$ up to the given order $N$, is in a
good accordance with the numeric simulations (\cite{JJS1999}) for
the ClOH molecule. However, with the vibrational levels more
strongly excited, one should take into account possible resonances
$p:q$ between the different oscillators of the molecular system. In
the second article, we complete the method of construction and
counting of coupling operators induced by the resonance for the case
of one single resonance $p:q$.

\section*{Acknowledgments}\label{Ack}
Authors thank Prof. Oleg N. Ulenikov and Elena S. Bekhtereva from
Tomsk Polytechnic University, for valuable discussions. Part of the work was 
supported by the project "Leading Russian Research
Universities" (grant FTI-120 of the Tomsk Polytechnic University).

\end{document}